\documentclass[submission,copyright,creativecommons]{eptcs}
\providecommand{\event}{ICLP 2020} 
\usepackage{underscore}           

\usepackage{BP-ICLP2020}
\usepackage{mathptmx}
\usepackage{graphicx}
\usepackage{url}
\usepackage{amssymb}
\usepackage{tikz}
\usepackage{comment}

\newtheorem{definition}{Definition}[section]
\newtheorem{notation}{Notation}[section]

\newtheorem{example}{Example}
\newcommand{\tuple}[1]{\langle #1 \rangle}
\newcommand{\lit}{\ensuremath{\phi}}

\newcommand{\BP}{\ensuremath{\textrm{BP}}}
\newcommand{\AR}{\mathcal{A}}
\newcommand{\defeat}{\leadsto}
\newcommand{\Conc}{\ensuremath{Conc}}
\newcommand{\Sub}{\ensuremath{Sub}}
\newcommand{\DirectSub}{\ensuremath{DirectSub}}

\newcommand{\TopRule}{\ensuremath{TopRule}}
\newcommand{\IN}{\mbox{\scalebox{0.75}{$\mathsf{IN}$}}}
\newcommand{\OUT}{\mbox{\scalebox{0.75}{$\mathsf{OUT}$}}}
\newcommand{\UND}{\mbox{\scalebox{0.75}{$\mathsf{UND}$}}}

\title{Burden of Persuasion in Argumentation}

\author{Roberta Calegari
\institute{CIRSFID -- Alma AI, \\University of Bologna, Italy}\thanks{Roberta Calegari and Giovanni Sartor have been supported by the ``CompuLaw'' project, funded by the European Research Council (ERC) under the European Union’s Horizon 2020 research and innovation programme (Grant Agreement No. 833647).}
\email{roberta.calegari@unibo.it}
\and
Giovanni Sartor
 \institute{CIRSFID -- Alma AI, \\University of Bologna, Italy}
\institute{European University Institute,  \\Florence, Italy}
\email{giovanni.sartor@unibo.it}
}
\def\titlerunning{Burden of Persuasion in Argumentation}
\def\authorrunning{R. Calegari and G. Sartor}
\begin{document}
\maketitle

\begin{abstract}
This paper provides a formal model for the burden of persuasion in dialogues, and in particular, in legal proceedings. 
The model shows how an allocation of the burden of persuasion may induce single outcomes in dialectical contexts in which, without such an allocation, the status of conflicting arguments would remain undecided.  Our approach is based on a two-stage labelling. The first-stage labelling determines what arguments are accepted, rejected  or undecided, regardless of the allocation of burden.
The second-stage labelling revises the dialectical status of first-stage undecided arguments, according to burdens of persuasion. The labelling is finally extended in such a way as to obtain a complete labelling.
Our model combines two ideas that have emerged in the debate on the burden of persuasion: the idea that the burden of persuasion determines the solution of conflicts between arguments, and the idea that its satisfaction depends on the dialectical status of the arguments concerned. Our approach also addresses inversions of the burden of persuasion, namely, cases in which the burden of persuasion over an argument does not extend to its subarguments.
\end{abstract}


\section{Introduction}\labelsec{section:introduction}
The concept of a burden of proof plays a key role in argumentation and in law. It is a complex and controversial idea for which no agreed theory exists \cite{walton2014}.
Generally, we can say that burdens of proof distribute dialectical responsibilities between the parties of argumentation: when a party has a burden of proof relative to a claim, that party will fail to establish the claim unless the party provides the kind of argument or evidence that is required to meet the burden.
Burdens of proof can complement dialectical frameworks for argumentation systems. In particular, burdens of proof are important in adversarial contexts: they facilitate reaching a single outcome in contexts of doubt and lack of information.  

In the legal domain, two burdens are distinguished: the burden of production (also called burden of providing evidence), and the burden of persuasion.

The burden of production concerns the necessity of providing evidence, in order to establish a claim. The evidence must be sufficient to establish the claim in the absence of counterarguments. When knowledge is represented through a set of rules and exceptions,  the party who is interested in establishing the conclusion of a rule has the burden of production relative to the strong (by default non-assumable) elements in the antecedent condition of the rule. The other party -- who is interested in preventing the derivation of the conclusion of the rule -- has the burden of production relative to the exceptions to the rule.
For instance, assume the law includes both the rule that there is murder if there are killing and intent, and the exception according to which there is no murder -- or the murder rule does not apply -- in case of self-defence. If evidence is not provided for both killing and intent, the prosecution will fail to substantiate the  murder claim; on the other hand, if no evidence is provided for self-defence, the defendant will fail to substantiate the claim that there is no murder, or that the murder rule is not to be applied.

The burden of persuasion concerns the necessity of providing a dialectically convincing argument, in order to establish a claim.  In order to be convincing, the argument must prevail over all counterarguments -- which are non-rejected on other grounds-- to an extent that is determined by the applicable standard of proof. If this is not the case, then the argument burdened with persuasion must be rejected. Not only will that argument fail to justify its conclusion, but it will also be unable to successfully challenge any other arguments.

For instance, in criminal cases, the burden of persuasion for the non-existence of self-defence is on the prosecution, even if the burden of production is on the defendant. Assume that it has been established that the defendant has committed the criminal action intentionally, while it remains doubtful whether she acted in self-defence since there are arguments for and against self-defence.
Under such circumstances, since prosecution has the burden of providing a convincing argument for the  non-existence of self-defence,  its doubtful argument against self-defence will be rejected. As consequence self-defence will be assumed, and the argument for murder will be rejected.

Our approach to the burden of persuasion expands the model proposed by Prakken and Sartor \cite{PrakkenSartor-bp-2010}, according to which  arguments for a claim burdened with persuasion will be strictly defeated unless they strictly prevail over arguments denying that claim. In our approach, to assess whether an argument for a claim $\phi$ meets the burden of persuasion, we not only consider whether the argument strictly prevails over arguments for $\bar\phi$. We also required that all (direct) subarguments are accepted.  This idea corresponds to the model of burdens in argumentation proposed by  Gordon et al. \cite{Gordon2007}, who assess the satisfaction of burdens of proofs based on the dialectical status of the concerned argument.

Our model takes into account that a multistep argument can be attacked not only by contradicting its final conclusion but also by challenging its earlier steps. Thus, for an argument to meet a burden of persuasion, there must be no doubt -- i.e., unsolved conflict -- also relative to such steps. For instance, assume that the claim that the defendant did not act is self-defence is supported by testimony, and that it is doubtful whether the witness is reliable. In such a case too, the prosecution will not have met the burden on non-self-defence, and the argument against self defence will be rejected.

Our model originates from legal considerations and is applied to legal examples and is mostly relevant for the development of AI tools to support legal argumentation \cite{logictech-information11}. 
However, the issue of the burden of proof has a significance that goes beyond the law; it involves other domains as well -- public discourse, risk management, etc. -- in which multiple agents, having different tasks and responsibilities,  provide evidence and arguments.

Formal models of argumentation are making significant and increasing contributions to AI---from defining the semantics of logic programs to implementing persuasive medical diagnostic systems, up to studying negotiation dialogues in multi-agent systems. The possibility to formalise burdens of proofs can be useful in multi-agent scenarios where agents have access to different pieces of information and may strategically withdraw information, and decisions are needed under situations of uncertainty. 

\smallskip

\textbf{Outline.} The paper is organised as follows, \xs{arg-framework} introduces a simple argumentation setting which is the basis of our model. \xs{running-example} introduces two running examples.
\xs{section:labelling-bp-definition} presents the contribution of the paper, namely, the concept of burden of persuasion in the labelling stage. Final remarks are provided in \xs{section:conclusion}.

\section{Background notion: argumentation framework language}\labelsec{arg-framework}
To illustrate our model of the burden of persuasion, we use a lightweight ASPIC\textsuperscript{+}-like system for structured  argumentation \cite{Prakken2010AF}. We do not include some key aspects of ASPIC\textsuperscript{+},  since they are not needed to introduce our model and present our examples: we do not distinguish between premises, assumptions and rules, or between strict and defeasible rules, or between rebutting and undercutting attack, or between different criteria for priority between arguments. Our characterisation of burdens of persuasions can, however, be applied to the full  ASPIC\textsuperscript{+}, or to other similar argumentation systems.

In our model, arguments are produced from a set of defeasible rules, and attacks between arguments are captured by argumentation graphs, to which a  labelling semantics is applied. 

\subsection{Defeasible theories and argumentation graphs} \labelssec{section:language}

A literal is an atomic proposition or the negation of an atomic proposition.

\begin{notation}
For any  literal $\phi$, its complement is written $\bar\phi$. That is, if $\phi$ is a proposition $p$ then $\bar\phi = \neg p$, while if $\phi$ is $\neg p$ then $\bar \phi$ is $p$.
\end{notation}

Literals are put into relation through defeasible rules. 

\begin{definition}\label{definition:def-rule}
A \textbf{defeasible rule} has the form: \begin{center}
r:\quad $\phi_1,...,\phi_n \Rightarrow \phi$
\end{center}
with $0 \leq n$, and where
\begin{itemize}
    \item $r$ is the unique identifier of the rule, and
    \item  each $\phi_1,\ldots \phi_n, \phi$ is a literal. 
\end{itemize}
\end{definition}

\noindent A superiority relation $\succ$ is defined over rules: $s \succ r$ states that rule $s$ prevails over rule $r$.
\begin{definition}\label{definition:sup-rel}
A \textbf{superiority relation} $\succ$ over a set of rules $Rules$ is an antireflexive and antisymmetric binary relation over $Rules$, i.e., $\succ \subseteq Rules \times Rules$.
\end{definition}

\noindent A defeasible theory consists in a set of rules and a superiority relation over the rules.
\begin{definition}\label{definition:def-theory}
A \textbf{defeasible theory} is a tuple $\langle Rules, \succ \rangle$ where $Rules$ is a set of rules, and $\succ$ is a superiority relation over $Rules$.
\end{definition}

Given a defeasible theory and by chaining rules from the theory, we can construct arguments, as specified in the following  definition, cf. \cite{DBLP:journals/argcom/ModgilP14,10.1016/j.artint.2007.02.003,DBLP:journals/ai/Vreeswijk97}.

\begin{definition}\label{def:argument}
An \textbf{argument} $\mathsf{A}$ constructed from a defeasible theory $\langle Rules, \succ \rangle$ is a finite construct of the form: 
\begin{center}
$\mathsf{A}: \, \mathsf{A}_1, \ldots \mathsf{A}_n \Rightarrow_r \lit$
\end{center}
with $0 \leq n$, where
\begin{itemize} 
\item $\mathsf{A}$ is the unique identifier of the argument;
\item $\mathsf{A}_1, \ldots, \mathsf{A}_n$ are arguments constructed from the defeasible theory $\langle Rules, \succ \rangle$;
\item $\lit$ is the \emph{conclusion} of the argument, denoted $\Conc(\mathsf{A})$;
\item  $r: \Conc(\mathsf{A}_1), \ldots, \Conc(\mathsf{A}_n) \Rightarrow \lit$ is the top rule of $\mathsf{A}$, denoted  $\TopRule(\mathsf{A})$.
\end{itemize}
\end{definition}

\begin{notation}
Given an argument $\mathsf{A}: \, \mathsf{A}_1, \ldots \mathsf{A}_n \Rightarrow_r \lit$ as in definition \ref{def:argument},
\begin{itemize}	
\item $\Sub(\mathsf{A})$ denotes the   \textbf{set of subarguments} of $\mathsf{A}$, i.e.,	$\Sub(\mathsf{A}) = \Sub(\mathsf{A}_1) \cup \ldots \cup \Sub(\mathsf{A}_n) \cup\{\mathsf{A}\}$;

\item $\DirectSub(\mathsf{A})$ denotes the \textbf{direct subarguments} of $\mathsf{A}$, i.e. $\DirectSub(\mathsf{A}) = \{\mathsf{A}_1, \ldots, \mathsf{A}_n \}$.
\end{itemize}
\end{notation}

We define the preference over arguments through  a  last-link ordering according to which an argument $\mathsf{A}$ is preferred over another argument $\mathsf{B}$ if the top rule of  $\mathsf{A}$ is stronger than the top rule of $\mathsf{B}$.

\begin{definition}
\label{definition:preference-arg}
A \textbf{preference relation} $\succ$ is a binary relation over a set of arguments  $\mathcal{A}$: an argument $\mathsf{A}$ is preferred to argument $B$, denoted $\mathsf{A} \succ \mathsf{B}$, iff  $\TopRule(\mathsf{A}) \succ \TopRule(\mathsf{B})$. 
\end{definition}

Argument $\mathsf{A}$ attacks argument $\mathsf{B}$ if $\mathsf{A}$'s conclusion is incompatible with the conclusion of a non-inferior subargument of $\mathsf{B}$.

\begin{definition}
\label{definition:attack} 
An \textbf{attack relation} $\defeat$ is a binary relation over a set of arguments  $\mathcal{A}$: for any $A, B\in  \mathcal{A}$, 
 $\mathsf{A}$ attacks  $\mathsf{B}$  iff  $\exists \mathsf{B}' \in \Sub(\mathsf{B})$ such that $\Conc(\mathsf{A}) = \overline{\Conc(\mathsf{B}')}$, and $\mathsf{B}' \not\succ \mathsf{A}$.
\end{definition}

Given the arguments and attacks obtained from a defeasible theory, we can create argumentation graphs.

\begin{definition}
\label{def:ag}
An \textbf{argumentation graph} constructed from a defeasible theory $T$  is a tuple $\tuple{\AR, \defeat}$ where $\AR$ is the set of all arguments constructed from $T$, and $\defeat$ is an \emph{attack} relation over $\AR$.
\end{definition}

\noindent In the following, we assume that all argumentation graphs are constructed from a defeasible theory.

\begin{notation}
Given an argumentation graph $G = \tuple{\AR,\defeat}$, we may write $\mathcal{A}_G$, and $\defeat_G$ to denote the graph's arguments, and attacks respectively.
\end{notation}

\begin{example}[Argumentation graph]\label{example:basic-graph}
Let us consider the following rules:
\begin{center}
\begin{tabular}{l l l l l}
 $r0:$  & $\Rightarrow p$           &\hspace{0.2cm} &$r1:$  &$\Rightarrow q$\\
 $r2:$  & $p\Rightarrow \neg r$     &               &$r3:$  &$q \Rightarrow r$\\
 $r4:$  & $r \Rightarrow s$           &               &$r5:$  &$\Rightarrow \neg s$ \\
\end{tabular}
\end{center}

\noindent with $r2 \succ r3$. Accordingly to the above definitions, we can then build  the following arguments: 
\begin{center}
\begin{tabular}{ l l l l l}
$\mathsf{A1}:\quad \Rightarrow p$ &\hspace{0.2cm} &  $\mathsf{B1}: \quad \Rightarrow q$ & \hspace{0.2cm} &  $\mathsf{C1}:\quad \Rightarrow \neg s$ \\
$\mathsf{A2}:\quad \mathsf{A1} \Rightarrow \neg r$
 & & $\mathsf{B2}: \quad \mathsf{1} \Rightarrow r$ \\
 & & $\mathsf{B3}:\quad \mathsf{B2}\Rightarrow s$ \\
\end{tabular}
\end{center}

\noindent Arguments $\mathsf{A1}$ and $\mathsf{B1}$ are subarguments of  $\mathsf{A2}$ and $\mathsf{B2}$ respectively.  $\mathsf{B2}$ is a subargument of $\mathsf{B3}$.  $\mathsf{A2}$ attacks $\mathsf{B2}$ and $\mathsf{B3}$,  while $\mathsf{C1}$ and $\mathsf{B3}$ attack each other. 
The superiority relation between $r2$ and $r3$ makes it possible to solve the conflict between arguments $\mathsf{B2}$ and $\mathsf{A2}$ in favour of $\mathsf{A2}$.
The corresponding argumentation graph is shown in \xf{basic-graph-fig}.

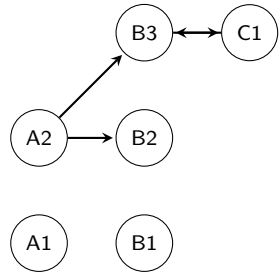
\begin{figure}[!ht]
\centering
\begin{tikzpicture}[node distance=1.4cm,main node/.style={circle,fill=white!20,draw,font=\sffamily\scriptsize}]

    \node[main node][label={[xshift=0.45cm, yshift=-0.9cm]}] (a5) {$\mathsf{B3}$};
    \node[main node][label={[xshift=-0.5cm, yshift=-0.5cm]}] (a3) [below of=a5] {$\mathsf{B2}$};
    \node[main node][label={[xshift=-0.5cm, yshift=-0.5cm]}] (a1) [below of=a3] {$\mathsf{B1}$};      
        
    \node[main node][label={[xshift=0.5cm, yshift=-0.5cm]}]  (a4) [left of=a3] {$\mathsf{A2}$};
    \node[main node][label={[xshift=-0.5cm, yshift=-0.5cm]}] (a0) [below of=a4] {$\mathsf{A1}$};

    \node[main node][label={[xshift=-0.6cm, yshift=-0.5cm]}] (a2) [right of=a5] {$\mathsf{C1}$};
    
    \path[->,>=stealth,shorten >=1pt,auto, thick,every node/.style={font=\sffamily\small}]
    (a2) edge node {} (a5)
    (a5) edge node {} (a2)
    (a4) edge node [right] {} (a5)
    (a4) edge node {} (a3);
    
\end{tikzpicture}
\caption{Argumentation graph corresponding to the theory of Example \ref{example:basic-graph}.}
\labelfig{basic-graph-fig}
\end{figure}

\hfill $\square$
\end{example}

\subsection{Labelling semantics}\labelssec{labelling-definition}
We use $\{\IN, \OUT, \UND\}$-labellings where any argument is associated with one label which is either $\IN$, $\OUT$, $\UND$, respectively meaning that the argument is accepted, rejected, or undecided.

\begin{definition}
A $\{\IN, \OUT, \UND \}$\textbf{-labelling} of an argumentation graph $G$ is a total function 
$$\AR_G \rightarrow  \{\IN, \OUT, \UND\}.$$
\end{definition}

\begin{notation}
Given a labelling $L$, we write $\IN(L)$ for $\{\mathsf{A} \mid L(\mathsf{A}) = \IN\}$, $\OUT(L)$ for $\{\mathsf{A} \mid L(\mathsf{A}) = \OUT\}$ and $\UND(L)$ for $\{\mathsf{A} \mid L(\mathsf{A}) = \UND\}$.
\end{notation}

\begin{definition}
\sloppy  An $\{\IN, \OUT, \UND\}$-labelling of an argumentation graph  $G$ is a \textbf{complete $\{\IN, \OUT, \UND\}$-labelling} iff for every argument $\mathsf{A}$ in $\mathcal{A}_G$:
\begin{itemize}
	\item[$\bullet$] $\mathsf{A}$ is  $\IN$ iff all attackers of $\mathsf{A}$ are $\OUT$,
	\item[$\bullet$] $\mathsf{A}$ is  $\OUT$ iff $\mathsf{A}$ has (at least) an attacker $\IN$.
\end{itemize}
\end{definition}

\noindent Since $\{\IN, \OUT, \UND\}$-labellings are total functions, if an argument is not  $\IN$ or $\OUT$, then it is $\UND$.

\sloppy An argumentation graph may have several complete $\{\IN, \OUT, \UND\}$-labellings, we will focus on the unique complete labelling with the smallest set of labels $\IN$, namely the grounded $\{\IN, \OUT, \UND\}$-labelling.

\begin{definition}\label{grounded-labelling-def}
\sloppy  A \textbf{grounded} $\{\IN, \OUT, \UND\}$-labelling $L$ of an argumentation graph $G$ is a complete $\{\IN, \OUT, \UND\}$-labelling of $G$ such that $\IN(L)$ is minimal.
\end{definition}

\begin{example}[Grounded $\{ \IN, \OUT, \UND \}$- labelling]
The grounded $\{ \IN, \OUT, \UND\}$-labelling of the argumentation graph of \xf{basic-graph-fig} is shown in \xf{basic-grounded-graph-fig}.
\begin{figure}[!ht]
\centering
\begin{tikzpicture}[node distance=1.4cm,main node/.style={circle,fill=white!20,draw,font=\sffamily\scriptsize}]
    
    \node[main node][label={[xshift=0.6cm, yshift=-0.3cm]\IN}] (a1) [fill = green!40] {$\mathsf{B1}$};
    \node[main node][label={[xshift=0.6cm, yshift=-0.3cm]\OUT}] (a3) [above of=a1, fill = red!40] {$\mathsf{B2}$};
    \node[main node][label={[xshift=-0.7cm, yshift=-0.5cm]\OUT}] (a5) [above of=a3, fill = red!40] {$\mathsf{B3}$};
    
    \node[main node][label={[xshift=-0.6cm, yshift=-0.3cm]\IN}] (a0) [left of=a1, fill = green!40] {$\mathsf{A1}$};
    \node[main node][label={[xshift=-0.6cm, yshift=-0.3cm]\IN}] (a4) [above of=a0, fill = green!40] {$\mathsf{A2}$};

    \node[main node][label={[xshift=0.6cm, yshift=-0.5cm]\IN}] (a2) [right of=a5, fill = green!40] {$\mathsf{C1}$};
    
    \path[->,>=stealth,shorten >=1pt,auto, thick,every node/.style={font=\sffamily\small}]
    (a2) edge node {} (a5)
    (a5) edge node {} (a2)
    (a4) edge node [right] {} (a5)
    (a4) edge node {} (a3);
    
\end{tikzpicture}
\caption{Grounded  {\scriptsize{$\{ \mathsf{IN}, \mathsf{OUT}, \mathsf{UND}\}$}}-labelling of the argumentation graph of 
\xf{basic-graph-fig}}
\labelfig{basic-grounded-graph-fig}
\end{figure}
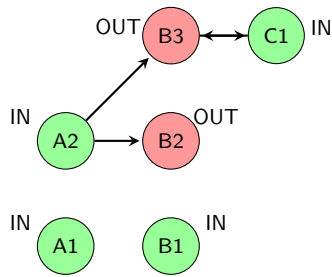

\hfill $\square$
\end{example}

\section{Running examples}\labelsec{running-example}
In this section, we introduce two examples concerning legal cases, a civil law one and a criminal law one. We shall here consider grounded labelling in the absence of burdens of persuasion and then, in the next section, the labelling obtained according to the applicable burdens. 

\begin{example}[Civil law example: medical malpractice]\label{example-medical-negligence}

Assume the applicable law according to which doctors are liable for the harm suffered by a patient, if they were negligent in treating the patient. 
However, there is an exception:  doctors can avoid liability if they show they were not negligent.  Assume also that doctors are considered to be non-negligent -- i.e., diligent -- if they followed the medical guidelines that govern the case. Assume that in the case at hand it is uncertain whether the doctor has followed the guidelines.   
Accordingly, let us consider the following rules:
\begin{center}
\begin{tabular}{l l l l l}
r1: &$\Rightarrow \neg guidelines\quad$ 		& &r2: &$\Rightarrow  guidelines$\\
r3: &$\neg guidelines \Rightarrow negligent$ 	& &r4: &$ guidelines \Rightarrow \neg negligent$\\
r5: &$\Rightarrow harm $ 			& &r6: &$\neg negligent \Rightarrow \neg liable$ \\
r7: & $harm \Rightarrow liable$ \\
\end{tabular}
\end{center}

\noindent with $r6 \succ r7$. We can then build the following arguments:
\begin{center}
\begin{tabular}{l l l l l}
$\mathsf{A1}:\quad \Rightarrow \neg guidelines$ & \hspace{0.2cm} & $\mathsf{B}1:\quad \Rightarrow guidelines$ & \hspace{0.2cm} & $\mathsf{C1}:\quad  \Rightarrow harm$\\
$\mathsf{A2}:\quad \mathsf{A1} \Rightarrow negligent$ & & $\mathsf{B2}:\quad \mathsf{B1}\Rightarrow \neg negligent$ & \hspace{0.2cm} & $\mathsf{C2}: \quad \mathsf{C1} \Rightarrow liable$ \\
 & &  $\mathsf{B3}:\quad \mathsf{B2} \Rightarrow \neg liable$\\
 
\end{tabular}
\end{center}

\noindent The argumentation graph and its grounded $\{ \IN, \OUT, \UND \}$-labelling are depicted in Figure \ref{medical-example-no-burden}, in which  all arguments are $\UND$, except argument $\mathsf{C1}$.

\begin{figure}[!ht]
\centering
\begin{tikzpicture}[node distance=1.4cm,main node/.style={circle,fill=white!20,draw,font=\sffamily\scriptsize}]
    \node[main node][label={[xshift=-0.65cm, yshift=-0.9cm]\UND}] (guidelines) [fill = blue!40] {$\mathsf{A1}$};
    \node[main node][label={[xshift=-0.65cm, yshift=-0.9cm]\UND}] (negligent) [above  of=guidelines, fill = blue!40] {$\mathsf{A2}$};
    
    \node[main node][label={[xshift=0.65cm, yshift=-0.9cm]\UND}] (noGuidelines) [right of=guidelines, fill = blue!40] {$\mathsf{B1}$};
    \node[main node][label={[xshift=0.65cm, yshift=-0.9cm]\UND}] (noNegligent) [above of=noGuidelines, fill = blue!40] {$\mathsf{B2}$};
    \node[main node][label={[xshift=0.65cm, yshift=-0.9cm]\UND}] (noLiable) [above of=noNegligent, fill = blue!40] {$\mathsf{B3}$};
    
    \node[main node][label={[xshift=0.65cm, yshift=-0.9cm]\IN}] (harm) [right of=noGuidelines, fill = green!40] {$\mathsf{C1}$};
    \node[main node][label={[xshift=0.65cm, yshift=-0.9cm]\UND}] (liable) [above  of=harm, fill = blue!40] {$\mathsf{C2}$}; 
     
    \path[->,shorten >=1pt,auto, thick,every node/.style={font=\sffamily\small}]
    (guidelines) edge node {} (noGuidelines)
    (noGuidelines) edge node [right] {} (guidelines)

    (guidelines) edge node {} (noNegligent)
    (noGuidelines) edge node [right] {} (negligent)

    (negligent) edge node {} (noNegligent)
    (noNegligent) edge node [right] {} (negligent)
    
    (guidelines) edge node {} (noLiable)
    
    (negligent) edge node {} (noLiable)
    
    (noLiable) edge node [right] {} (liable);   
\end{tikzpicture}
\caption{Grounded \scriptsize{$\{ \mathsf{IN}, \mathsf{OUT}, \mathsf{UND}\}$}-labelling of Example \ref{example-medical-negligence} in the absence of burdens of persuasion.}
\label{medical-example-no-burden}
\end{figure}
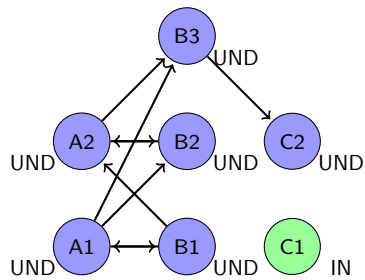

The result shown in Figure \ref{medical-example-no-burden} is not satisfactory, according to the law, since it does not take into account the applicable burdens of persuasion. The doctor should have lost the case -- i.e., be declared liable -- since she failed to discharge her burden of proving  she was non-negligent---namely, she didn't satisfy her burden of persuasion for her non-negligence. The doctor's failure results from the fact that it remains uncertain whether she followed the guidelines. To capture this aspect of the argument, we need to provide a  model for the burden of persuasion.

\hfill $\square$
\end{example}

\begin{example}[Criminal law example: self-defence in murder case]\label{example-criminal}
Let us consider a case in which a woman shot a robber in her house, killing him. The applicable law includes the rule according to which intentional killing constitutes murder, as well as the exception according to which there is no murder if the victim was killed in self-defence.
Assume that it has been established that the woman shot the robber and that she did so intentionally, i.e, the prosecution has met her burden of persuasion with regard to both killing and intent. 
However, it remains uncertain, due to conflicting evidence, whether the robber was threatening her with a gun, as claimed by the defence, or was running away after having been discovered, as claimed by the prosecution.
Accordingly, let us consider the following rules:

\begin{center}
\begin{tabular}{l l l l l}
r1:& $\Rightarrow killed$                        & & r2:& $\Rightarrow intention$ \\ 
r3:& $\Rightarrow threatWithWeapon\qquad$        & & r4:& $\Rightarrow \neg threatWithWeapon$\\
r5:& $threatWithWeapon \Rightarrow selfDefence$  & & r6:& $\neg threatWithWeapon \Rightarrow \neg selfDefence$\\
r7:& $selfDefence \Rightarrow \neg murder$       & & r8:& $killed$, $intention \Rightarrow murder$\\
\end{tabular}
\end{center}

\noindent with  $r7 \succ r8$. We can build the following arguments:
\begin{center}
\begin{tabular}{l l l l l l l l}
$\mathsf{A1}:$& $\Rightarrow killed$          & &$\mathsf{B1}:$& $\Rightarrow threatWithWeapon$  & &$\mathsf{C1}:$& $\Rightarrow  \neg threatWithWeapon$   \\
$\mathsf{A2}:$& $\Rightarrow intention$ 	  & &$\mathsf{B2}:$& $\mathsf{B1} \Rightarrow selfDefence$ & &$\mathsf{C2}:$& $\mathsf{C1} \Rightarrow \neg selfDefence$ \\
$\mathsf{A3}:$ & $\mathsf{A1}, \mathsf{A2} \Rightarrow murder$             & & $\mathsf{B3}:$& $\mathsf{B2} \Rightarrow \neg murder$\\ 
\end{tabular}
\end{center}

\noindent In the grounded $\{ \IN, \OUT, \UND \}$-labelling of  \xf{criminal-example-no-burden}, all arguments are $\UND$, except for the undisputed facts.

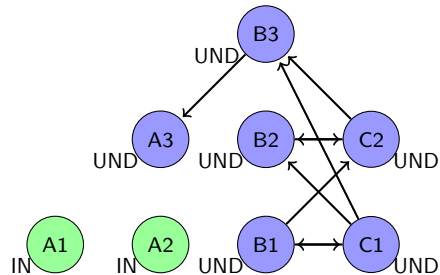
\begin{figure}[!ht]
\centering
\begin{tikzpicture}[node distance=1.4cm,main node/.style={circle,fill=white!20,draw,font=\sffamily\scriptsize}]
  \node[main node][label={[xshift=-0.6cm, yshift=-0.9cm]\UND}] (b1) [fill = blue!40] {$\mathsf{B1}$};
   \node[main node][label={[xshift=-0.6cm, yshift=-0.9cm]\UND}] (b2) [above of=b1, fill = blue!40] {$\mathsf{B2}$};
   \node[main node][label={[xshift=-0.65cm, yshift=-0.9cm]\UND}] (b3) [above of=b2, fill = blue!40] {$\mathsf{B3}$};

   \node[main node][label={[xshift=0.6cm, yshift=-0.9cm]\UND}] (c1) [right of=b1, fill = blue!40] {$\mathsf{C1}$};
  \node[main node][label={[xshift=0.6cm, yshift=-0.9cm]\UND}] (c2) [above of=c1, fill = blue!40] {$\mathsf{C2}$};
 
  \node[main node][label={[xshift=-0.45cm, yshift=-0.9cm]\IN}] (a2) [left of=b1, fill = green!40] {$\mathsf{A2}$};
  \node[main node][label={[xshift=-0.45cm, yshift=-0.9cm]\IN}] (a1) [left of=a2, fill = green!40] {$\mathsf{A1}$};
  \node[main node][label={[xshift=-0.6cm, yshift=-0.9cm ]\UND}] (a3)[left of=b2, fill = blue!40] {$\mathsf{A3}$};

    \path[->,shorten >=1pt,auto, thick,every node/.style={font=\sffamily\small}]
    (b1) edge node {} (c1) 
    (c1) edge node [right] {} (b1)
    (c1) edge node {} (b2)
    (c1) edge node {} (b3)
    
    (b1) edge node {} (c2)

    (b3) edge node [right] {} (a3)
    
    (b2) edge node {} (c2)
    (c2) edge node [right] {} (b2)
    (c2) edge node {} (b3);

\end{tikzpicture}
\caption{Grounded  \scriptsize{$\{ \mathsf{IN}, \mathsf{OUT}, \mathsf{UND}\}$}-labelling of Example \ref{example-criminal} in the absence of burdens of persuasion.
}
\labelfig{criminal-example-no-burden}
\end{figure}

Also in this case, the  labelling does not provide the legally correct answers, namely, acquittal. 
Acquittal results from the fact that the burden of persuasion falls on the non-existence of self-defence.  Prosecution failed to provide a convincing argument for that, since it remains dubious whether the woman who shot the robber was threatened with a weapon by the latter.
\end{example}

\section{Labelling and burden of persuasion}\labelsec{section:labelling-bp-definition}

Now, let us introduce the concept of burden of persuasion in a second labelling stage.

\begin{notation}
To indicate that there is a burden of persuasion on  every literal in a set of literals  $\Phi$, we write BP($\Phi$). If $\Phi$ is a singleton $\{\lit \}$,  we may simply write  BP($\lit$). 
\end{notation}

 
The burden of persuasion may be captured in subsequent labelling stages, cf. multi-labelling systems \cite{DBLP:journals/jair/BaroniR19}. A burden of persuasion labelling (BP-labelling) based on an $\{\IN, \OUT, \UND \}$-labelling  assigns $\OUT^*$ and $\IN^*$ labels to $\UND$ arguments depending on the allocation of the burden of persuasion. $\IN$ and $\OUT$ arguments maintain their status, i.e., are labelled $\IN^*$ and $\OUT^*$ respectively.

\subsection{BP-labelling definition}

\begin{definition}
A \textbf{$\{\IN^*, \OUT^*, \UND^*\}$-labelling} of an argumentation graph $G$ is a total function $\AR_G \rightarrow  \{\IN^*, \OUT^*, \UND^*\}$.
\end{definition}

\begin{definition}\label{burdenLabelling}
Let $G$ be  an argumentation graph,  $L$ the grounded $\{\IN, \OUT, \UND \}$-labelling of $G$, and $\Phi$  a consistent set of literals.
A \textbf{BP-labelling} of $G$, relative to burdens of persuasion $\BP(\Phi$), is a $\{\IN^{*}, \OUT^{*}, \UND^{*}\}$-labelling such that $\forall (\mathsf{A}: \, \mathsf{A}_1, \ldots \mathsf{A}_n \Rightarrow_r \lit) \in \mathcal{A}_G$

\begin{enumerate}

    \item if $\mathsf{A} \in \IN(L)$ then $\mathsf{A}\in \IN^{*}$;
    \item if $\mathsf{A} \in \OUT(L)$ then $\mathsf{A} \in \OUT^{*}$;

    \item if $\mathsf{A} \in \UND(L)$ then
    \begin{enumerate}
        \item  $\mathsf{A} \in \IN^{*}$ iff 
            \begin{enumerate}
                \item $\bar \lit \in \Phi$ and 
                		\begin{itemize}
                		\item $\nexists$  $\mathsf{B} \in \mathcal{A}_G$ such that $\Conc(\mathsf{B})=\bar\lit$, $\mathsf{B}\succ \mathsf{A}$  and $\mathsf{B}$ is  $\IN^{*}$\\
			and   
			\item $\nexists{} \; \mathsf{A}' \in \DirectSub(\mathsf{A}): \mathsf{A}'$ is  $\OUT^{*}$
			\end{itemize}
            or 
                \item 
                $\bar \lit \not\in \Phi$ and 
                		\begin{itemize}
                		\item $\forall\;\mathsf{B} \in \mathcal{A}_G$ such that $\Conc(\mathsf{B})=\bar \lit$, $\mathsf{A} \not\succ \mathsf{B}:$ $\mathsf{B}$ is  $\OUT^{*}$
                		
			and 
                         \item $\forall \; \mathsf{A}' \in \DirectSub(\mathsf{A}): \mathsf{A}'$ is $\IN^{*}$;
                         \end{itemize}
            \end{enumerate}
        \item $\mathsf{A}\in\OUT^{*}$ iff  
        \begin{enumerate}
            \item $\lit \in \Phi$ and 
            	\begin{itemize}
                		\item $\exists\;\mathsf{B} \in \mathcal{A}_G$ such that $\Conc(\mathsf{B})=\bar \lit$, $\mathsf{A}\not\succ \mathsf{B}$ and $\mathsf{B}$ is not $\OUT^*$  \\
			or 
			\item $\exists \; \mathsf{A}' \in \DirectSub(\mathsf{A}): \mathsf{A}'$ is not  $\IN^{*}$ 
		\end{itemize}
	    or
            \item $\lit \not\in \Phi$ and  
            	\begin{itemize}
                		\item $\exists\;\mathsf{B} \in \mathcal{A}_G$ such that $\Conc(\mathsf{B})=\bar \lit$,  $\mathsf{B}\succ \mathsf{A}$ and $\mathsf{B}$ is  $\IN^*$  \\
			or 
			\item $\exists \; \mathsf{A}' \in \DirectSub(\mathsf{A}): \mathsf{A}'$ is  $\OUT^{*}$;
		\end{itemize}
        \end{enumerate}
    \item $\mathsf{A}\in\UND^{*}$ otherwise.
    \end{enumerate}
\end{enumerate}
\end{definition}
\noindent In Definition \ref{burdenLabelling}, items \emph{3.(a)} and \emph{3.(b)} concern respectively  conditions for acceptance and rejection based on burdens of persuasion.

\paragraph{Condition for acceptance.}
\begin{itemize}
\item Item \emph{3.(a)(i)} concerns the case in which the burden of persuasion in on the complement $\bar\lit$  of the conclusion $\lit$ of argument $\mathsf{A}$. A  counterargument   $\mathsf{B}$ for $\bar\lit$ is disfavoured by the burden of persuasion while $\mathsf{A}$  is favoured, i.e.,  $\mathsf{A}$ is to be accepted under conditions of uncertainty. Thus, acceptance of  $\mathsf{A}$ is not affected by $\mathsf{B}$  unless $\mathsf{B}$ is strongly superior to $\mathsf{A}$. and acceptance of  $\mathsf{A}$ is also not affected by the fact that a strict subarguments of $\mathsf{A}$ is $\UND^*$, rather than $\IN^*$.

\item Item \emph{3.(a)(ii)} concerns the case in which the conclusion of argument $\mathsf{A}$ is contradicted by a counterargument $\mathsf{B}$ upon which there is no burden of persuasion. Then there is no favour for $\mathsf{A}$. Acceptance of   $\mathsf{A}$ may be affected by $\mathsf{B}$ whenever $\mathsf{A}$ is not strictly superior to $B$, and acceptance of   $\mathsf{A}$ is also affected by the fact that one $\mathsf{A}$'s strict subarguments is $\UND^*$.
\end{itemize}

\paragraph{Condition for rejection.}
\begin{itemize}
\item Item \emph{3.(b)(i)} concerns the case in which the burden of persuasion is on the conclusion of argument $\mathsf{A}$, so that  $\mathsf{A}$ is disfavoured by the burden of persuasion. Then the rejection of $\mathsf{A}$ may be determined by a counterargument $\mathsf{B}$ that is uncertain ($\UND^*$)  and also  by any  uncertainty on one of $\mathsf{A}$'s strict subarguments.

\item  Item \emph{3.(b)(i)} concerns the case in which there is no burden of persuasion on the conclusion of argument $\mathsf{A}$. 
Then the rejection of $\mathsf{A}$  is only  determined by a counterargument $\mathsf{B}$ which   $\mathsf{A}$ which is $\IN^*$ or by a strict subargument of  $\mathsf{A}$ that is $\OUT^*$. 
\end{itemize}

\subsection{Completion of a BP-labellings}
A BP-labelling may not be complete, in the sense that an argument favoured by the burden of persuasion may be $\IN^*$ even is some of its subarguments are $\UND^*$. As shown in Example \ref{example-medical-negligence-bp} (\xf{bp-medical-malpractice-fig} left), an argument (e.g. $\mathsf{A2}$) may be $\IN^*$ because the burden of persuasion is on the complement of its conclusion, even though some of its subarguments are $\UND^*$. 
This is admitted by Definition \ref{burdenLabelling}, 3.(a)(i). The intuition  is the following:  when we have accepted a conclusion $\phi$ since the burden of persuasion is on $\bar\phi$, we may still remain uncertain on the premises for that conclusion. For instance, we may accept that the doctor was negligent since she failed to provide a convincing argument about why she was not negligent, and still be uncertain whether the doctor did not comply with the guidelines. 

The opposite approach is also possible: if we accept an argument based on the burden of persuasion, we are also bound to accept all of its subarguments. This approach leads to the concepts of completion and  grounding of BP-labellings.

\begin{definition}
A \textbf{completion $L_c$ of a BP-labelling $L$} of an argumentation graph $G$, relative to burdens $\BP(\Phi)$, is an $\{\IN^{*}, \OUT^{*}, \UND^{*}\}$-labelling such that   
\begin{enumerate}
    \item $\IN^{*}(L) \subseteq  \IN^{*}(L_c)$, and
    \item $\OUT^{*}(L) \subseteq  \OUT^{*}(L_c)$, and
    \item $\forall \mathsf{A}\in \mathcal{A}_G$
    \begin{enumerate}
   \item    $ \mathsf{A} \in \IN^{*}(L_c)$ iff all attackers of $\mathsf{A}$ are in $\OUT^{*}(L_c)$, and
   \item $\mathsf{A}\in \OUT^{*}(L_c)$ iff
        at least one attacker of $\mathsf{A}$ is $\IN^{*}(L_c)$.
    \end{enumerate}   
  \end{enumerate}
\end{definition}

\begin{definition}\label{groundeBurdenLabelling}
The \textbf{grounding $L_g$ of a   BP-labelling} $L$ of an argumentation graph $G$,  relative to burdens $\BP(\Phi)$, is a completion $L_c$ of $L$   such that
$\IN^{*}(L_c)$ is minimal.
\end{definition}

\subsection{Discussion of the examples}
Let us now apply these  definitions to our running examples, and to some further cases.

\begin{example}[Burden of persuasion on civil law example: medical malpractice]\label{example-medical-negligence-bp}
Continuing Example \ref{example-medical-negligence} presented in \xs{running-example}, assume -- accordingly to Italian Law -- to have $\BP(\neg negligent)$ (i.e., the doctor has to provide a convincing argument that she was non-negligent, i.e., that she was  diligent). 

By Definition  \ref{burdenLabelling} argument $\mathsf{A2}$  for the doctor's negligence must be $\IN^*$ and the argument for non-negligence is $\OUT^*$, but it  remains uncertain whether the doctor followed the guidelines (the corresponding arguments $\mathsf{A1}$ and $\mathsf{B1}$ are $\UND^*$). 

A different result can be obtained by using the grounding of the  BP-labelling. Then argument $\mathsf{A1}$ for $\neg guidelines$ is $\IN^*$ and argument $\mathsf{B1}$ for $guidelines$ is $\OUT^*$.
The BP-labelling is shown in \xf{bp-medical-malpractice-fig} (left) and the grounded labelling in \xf{bp-medical-malpractice-fig} (right).

\begin{figure}[!ht]
\centering
\begin{minipage}[c]{0.49\linewidth}
\centering
\begin{tikzpicture}[node distance=1.4cm,main node/.style={circle,fill=white!20,draw,font=\sffamily\scriptsize}]
    \node[main node][label={[xshift=-0.75cm, yshift=-0.9cm]$\UND^*$}] (guidelines) [fill = blue!40] {$\mathsf{A1}$};
    \node[main node][label={[xshift=-0.65cm, yshift=-0.99cm]$\IN^*$}] (negligent) [above of=guidelines, fill = green!40] {$\mathsf{A2}$};

    \node[main node][label={[xshift=0.65cm, yshift=-0.99cm]$\UND^*$}] (noGuidelines) [right of=guidelines, fill = blue!40] {$\mathsf{B1}$};
    \node[main node][label={[xshift=0.65cm, yshift=-0.99cm]$\OUT^*$}] (noNegligent) [above of=noGuidelines, fill = red!40] {$\mathsf{B2}$};
    \node[main node][label={[xshift=0.75cm, yshift=-0.9cm]$\OUT^*$}] (noLiable) [above of=noNegligent, fill = red!40] {$\mathsf{B3}$};
    
    \node[main node][label={[xshift=0.65cm, yshift=-0.9cm]$\IN^*$}] (harm) [right of=noGuidelines, fill = green!40] {$\mathsf{C1}$};
    \node[main node][label={[xshift=0.65cm, yshift=-0.9cm]$\IN^*$}] (liable) [above  of=harm, fill = green!40] {$\mathsf{C2}$};

    \path[->,shorten >=1pt,auto, thick,every node/.style={font=\sffamily\small}]
    (guidelines) edge node {} (noGuidelines)
    (noGuidelines) edge node [right] {} (guidelines)

    (guidelines) edge node {} (noNegligent)
    (noGuidelines) edge node [right] {} (negligent)

    (negligent) edge node {} (noNegligent)
    (noNegligent) edge node [right] {} (negligent)
    
    (guidelines) edge node {} (noLiable)
    
    (negligent) edge node {} (noLiable)
    
    (noLiable) edge node [right] {} (liable);   
    
\end{tikzpicture}
\end{minipage}
\begin{minipage}[c]{0.49\linewidth}
\centering
\begin{tikzpicture}[node distance=1.4cm,main node/.style={circle,fill=white!20,draw,font=\sffamily\scriptsize}]
    \node[main node][label={[xshift=-0.75cm, yshift=-0.9cm]$\IN^*$}] (guidelines) [fill = green!40] {$\mathsf{A1}$};
    \node[main node][label={[xshift=-0.75cm, yshift=-0.9cm]$\IN^*$}] (negligent) [above of=guidelines, fill = green!40] {$\mathsf{A2}$};

    \node[main node][label={[xshift=0.65cm, yshift=-0.99cm]$\OUT^*$}] (noGuidelines) [right of=guidelines, fill = red!40] {$\mathsf{B1}$};
    \node[main node][label={[xshift=0.65cm, yshift=-0.99cm]$\OUT^*$}] (noNegligent) [above of=noGuidelines, fill = red!40] {$\mathsf{B2}$};
    \node[main node][label={[xshift=0.75cm, yshift=-0.9cm]$\OUT^*$}] (noLiable) [above of=noNegligent, fill = red!40] {$\mathsf{B3}$};
    
    \node[main node][label={[xshift=0.65cm, yshift=-0.9cm]$\IN^*$}] (harm) [right of=noGuidelines, fill = green!40] {$\mathsf{C1}$};
    \node[main node][label={[xshift=0.65cm, yshift=-0.9cm]$\IN^*$}] (liable) [above  of=harm, fill = green!40] {$\mathsf{C2}$};

    \path[->,shorten >=1pt,auto, thick,every node/.style={font=\sffamily\small}]
    (guidelines) edge node {} (noGuidelines)
    (noGuidelines) edge node [right] {} (guidelines)

    (guidelines) edge node {} (noNegligent)
    (noGuidelines) edge node [right] {} (negligent)

    (negligent) edge node {} (noNegligent)
    (noNegligent) edge node [right] {} (negligent)
    
    (guidelines) edge node {} (noLiable)
    
    (negligent) edge node {} (noLiable)
    
    (noLiable) edge node [right] {} (liable);   
\end{tikzpicture}
\end{minipage}
\caption{BP-labelling (left) its grounding (right), with $\BP(\neg negligent)$, where  $\neg negligent = \Conc(\mathsf{B2})$.}
\labelfig{bp-medical-malpractice-fig}
\end{figure}
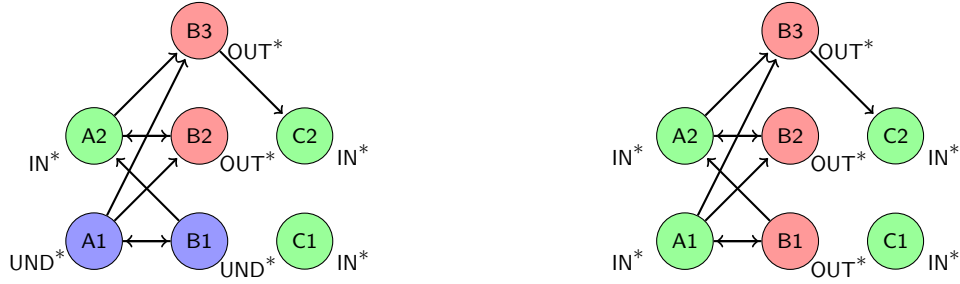
\hfill $\square$
\end{example}

\begin{example}[Burden of persuasion on criminal law example: self-defence in murder case]\label{example-criminal-bp}
In the criminal law example, the burden of persuasion is on the prosecution, both for the constitutive elements of the crime and for the non-existence of justifications (like self-defence). Thus, we specify  that  there is a burden of persuasion on $killing$ and $intent$ as well as on  the absence of self-defence, i.e., $BP(\{killed, intention, \neg selfDefence \})$. Hence, the prosecution has to persuade the judge that there was killing, that it was intentional, and that the killer did not act in self-defence. 
The grounding  is depicted in \xf{f-exampleCriminal-bp}. 

Note that while arguments $\mathsf{B1}$ and $\mathsf{C1}$ are $\UND^*$ in the BP-labelling,  in the grounding (\xf{f-exampleCriminal-bp} right),  $\mathsf{B1}$ is now labelled $\IN^{*}$ while $\mathsf{C1}$ is labelled $\OUT^{*}$.

\begin{figure}[!ht]
\centering
\begin{minipage}[c]{0.49\linewidth}
\centering
\begin{tikzpicture}[node distance=1.4cm,main node/.style={circle,fill=white!20,draw,font=\sffamily\scriptsize}]
  \node[main node][label={[xshift=-0.7cm, yshift=-0.9cm]$\UND^*$}] (b1) [fill = blue!40] {$\mathsf{B1}$};
   \node[main node][label={[xshift=-0.55cm, yshift=-0.9cm]$\IN^*$}] (b2) [above of=b1, fill = green!40] {$\mathsf{B2}$};
   \node[main node][label={[xshift=-0.55cm, yshift=-0.9cm]$\IN^*$}] (b3) [above of=b2, fill = green!40] {$\mathsf{B3}$};

   \node[main node][label={[xshift=0.75cm, yshift=-0.9cm]$\UND^*$}] (c1) [right of=b1, fill = blue!40] {$\mathsf{C1}$};
  \node[main node][label={[xshift=0.75cm, yshift=-0.9cm]$\OUT^*$}] (c2) [above of=c1, fill = red!40] {$\mathsf{C2}$};
 
   \node[main node][label={[xshift=-0.55cm, yshift=-0.9cm]$\IN^*$}] (a1) [left of=a2, fill = green!40] {$\mathsf{A1}$};
  \node[main node][label={[xshift=-0.55cm, yshift=-0.9cm]$\IN^*$}] (a2) [left of=b1, fill = green!40] {$\mathsf{A2}$};
  \node[main node][label={[xshift=-0.75cm, yshift=-0.9cm ]$\OUT^*$}] (a3)[above of=a2, fill = red!40] {$\mathsf{A3}$};

    \path[->,shorten >=1pt,auto, thick,every node/.style={font=\sffamily\small}]
    (b1) edge node {} (c1) 
    (c1) edge node [right] {} (b1)
    (c1) edge node {} (b2)
    (c1) edge node {} (b3)
    
    (b1) edge node {} (c2)

    (b3) edge node [right] {} (a3)
    
    (b2) edge node {} (c2)
    (c2) edge node [right] {} (b2)
    (c2) edge node {} (b3);

\end{tikzpicture}
\end{minipage}
\begin{minipage}[c]{0.49\linewidth}
\centering
\begin{tikzpicture}[node distance=1.4cm,main node/.style={circle,fill=white!20,draw,font=\sffamily\scriptsize}]
  \node[main node][label={[xshift=-0.55cm, yshift=-0.9cm]$\IN^*$}] (a1) [fill = green!40] {$\mathsf{B1}$};
   \node[main node][label={[xshift=-0.55cm, yshift=-0.9cm]$\IN^*$}] (a2) [above of=a1, fill = green!40] {$\mathsf{B2}$};
   \node[main node][label={[xshift=-0.55cm, yshift=-0.8cm]$\IN^*$}] (a3) [above of=a2, fill = green!40] {$\mathsf{B3}$};

   \node[main node][label={[xshift=0.75cm, yshift=-0.9cm]$\OUT^*$}] (b1) [right of=a1, fill = red!40] {$\mathsf{C1}$};
  \node[main node][label={[xshift=0.75cm, yshift=-0.9cm]$\OUT^*$}] (b2) [above of=b1, fill = red!40] {$\mathsf{C2}$};
 
   \node[main node][label={[xshift=-0.55cm, yshift=-0.9cm]$\IN^*$}] (c1) [left of=a1, fill = green!40] {$\mathsf{A2}$};
   \node[main node][label={[xshift=-0.55cm, yshift=-0.9cm]$\IN^*$}] (c2) [left of=c1, fill = green!40] {$\mathsf{A1}$};
  \node[main node][label={[xshift=-0.75cm, yshift=-0.9cm ]$\OUT^*$}] (c3)[above of=c1, fill = red!40] {$\mathsf{A3}$};

    \path[->,shorten >=1pt,auto, thick,every node/.style={font=\sffamily\small}]
    (a1) edge node {} (b1) 
    (b1) edge node [right] {} (a1)
    
    (a1) edge node {} (b2)
    (b1) edge node {} (a2)
    
    (a2) edge node {} (b2) 
    (b2) edge node {} (a3) 
    (b1) edge node {} (a3) 
    (b2) edge node [right] {} (a2)
    
    (a3) edge node {} (c3);

\end{tikzpicture}
\end{minipage}
\caption{BP-labelling (left) and its grounding (right), with the burden of persuasion $BP(\{ killed, intention, \neg selfDefence \})$.}
\labelfig{f-exampleCriminal-bp}
\end{figure}
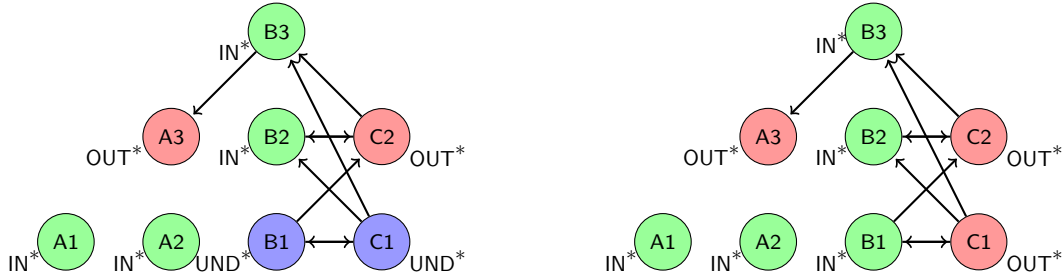
The BP-labelling provides the legally correct answer.  Prosecution failed to meet its  burden of proving non-self-defence (its argument $\mathsf{C2}$ is not convincing). Therefore, the argument for self-defence ($\mathsf{B2}$) is $\IN^*$ and so is the argument for not-murder ($\mathsf{B3}$), which means that the killer is to be acquitted.
\end{example}

\begin{example}[Conflicting burdens of persuasion]\label{example-multiple-bp}
When there are multiple  conflicts amongst arguments and subarguments, burdens disfavouring the conclusions of subarguments  are evaluated first. As we shall see in the next example, this allows us to capture inversions in the burden of persuasion.

Let us assume the following premises and rules with $BP(\{\neg a,b\})$:
\begin{center}
\begin{tabular}{l l }
$r1:\quad\Rightarrow a \quad$ & $r2:\quad a \Rightarrow b$\\
$r3:\quad \Rightarrow \neg a \quad$ &  $r4:\quad \neg a \Rightarrow \neg b$\\

\end{tabular}
\end{center}

We can then build the following arguments:
\begin{center}
\begin{tabular}{ l l l}
$\mathsf{A1}:\quad \Rightarrow a$ & & $\mathsf{B1}:\quad \Rightarrow \neg a$\\
$\mathsf{A2}:\quad \mathsf{A1} \Rightarrow b$ & & $\mathsf{B2}:\quad \mathsf{B1}\Rightarrow \neg b$\\
\end{tabular}
\end{center}

The corresponding grounding is depicted in \xf{example-multiple-bp-fig}. By Definition \ref{burdenLabelling}, $\mathsf{A1}$ and $\mathsf{A2}$ are $\IN^*$, while $\mathsf{B1}$ and $\mathsf{B2}$ are $\OUT^*$.

\begin{figure}[!ht]
\centering
\begin{tikzpicture}[node distance=1.4cm,main node/.style={circle,fill=white!20,draw,font=\sffamily\scriptsize}]
    \node[main node][label={[xshift=-0.55cm, yshift=-0.9cm]$\IN^*$}] (a) [fill = green!40] {$\mathsf{A1}$};
    \node[main node][label={[xshift=0.75cm, yshift=-0.9cm]$\OUT^*$}] (nega) [right of=guidelines, fill = red!40] {$\mathsf{B1}$};
    \node[main node][label={[xshift=-0.55cm, yshift=-0.9cm]$\IN^*$}] (b) [above of=guidelines, fill = green!40] {$\mathsf{A2}$};
    \node[main node][label={[xshift=0.75cm, yshift=-0.9cm]$\OUT^*$}] (negb) [above of=noGuidelines, fill = red!40] {$\mathsf{B}$};

    \path[->,shorten >=1pt,auto, thick,every node/.style={font=\sffamily\small}]
    (a) edge node {} (nega)
    (nega) edge node [right] {} (a)

    (b) edge node {} (negb)
    (negb) edge node [right] {} (b)

    (a) edge node {} (negb)
    (nega) edge node {} (b);   
    
\end{tikzpicture}
\caption{BP-labelling with $BP(\{\neg a, b\})$,where $\neg a = \Conc(\mathsf{B1})$ and $b=\Conc(\mathsf{A2})$.}
\labelfig{example-multiple-bp-fig}
\end{figure}
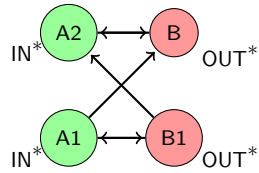

\end{example}

\begin{example}[Inversion of burden of persuasion: medical malpractice]\label{example-medical-negligence-double-bp}
 In legal and other contexts, often there is an inversion of the burden of persuasion: a burden of persuasion on the conclusion of an argument $\mathsf{A}$ coexists with a burden of persuasion on the complement of the conclusion of a subargument of $\mathsf{A}$. Developing Example \ref{example-medical-negligence-bp}, let us now  assume the following: the patient has the burden of persuasion on liability, while the doctor has the burden of persuasion on being non-negligent, i.e., $BP(\{liable, \neg negligent\})$ (negligence being a conclusion of a subargument of the doctor). 
Accordingly, the patient will lose if he fails to provide a convincing argument for the doctor's liability, which includes providing arguments for all preconditions of such liability. On the latter there is an inversion: it is up to the doctor to provide a convincing argument for her non-negligence.

From Definition \ref{burdenLabelling} it follows that:
\begin{itemize}
	\item $\mathsf{A2}$ and $\mathsf{C2}$ (concluding  $negligent$ and $liable$ respectively) are $\IN^*$, and 
	\item $\mathsf{B2}$ and $\mathsf{B3}$ (concluding $\neg negligent$ and $\neg liable$ respectively) are $\OUT^*$. 
\end{itemize}
Completing the labelling we also obtain that $\mathsf{A1}$ is $\IN^*$ and $\mathsf{B1}$ is $\OUT^*$.

The example shows that, according to Definition  \ref{burdenLabelling} the burdens of production are applied step by step: the burden are applied first to the assessment of subarguments, and then to the assessment of  arguments including them. Since the doctor could not provide a convincing argument for $\neg negligent$ (where she has the burden of persuasion), negligence is established. Therefore the doctor is unable to challenge the patient's liability through an argument based on her non-negligence.

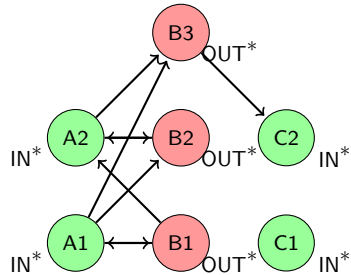
\begin{figure}[!ht]
\centering
\begin{tikzpicture}[node distance=1.4cm,main node/.style={circle,fill=white!20,draw,font=\sffamily\scriptsize}]
    \node[main node][label={[xshift=-0.65cm, yshift=-0.9cm]$\IN^*$}] (guidelines) [fill = green!40] {$\mathsf{A1}$};
    \node[main node][label={[xshift=0.65cm, yshift=-0.9cm]$\OUT^*$}] (noGuidelines) [right of=guidelines, fill = red!40] {$\mathsf{B1}$};
    \node[main node][label={[xshift=-0.65cm, yshift=-0.9cm]$\IN^*$}] (negligent) [above  of=guidelines, fill = green!40] {$\mathsf{A2}$};
    \node[main node][label={[xshift=0.65cm, yshift=-0.9cm]$\OUT^*$}] (noNegligent) [above of=noGuidelines, fill = red!40] {$\mathsf{B2}$};
    \node[main node][label={[xshift=0.65cm, yshift=-0.9cm]$\OUT^*$}] (noLiable) [above of=noNegligent, fill = red!40] {$\mathsf{B3}$};

   \node[main node][label={[xshift=0.65cm, yshift=-0.9cm]$\IN^*$}] (harm) [right of=noGuidelines, fill = green!40] {$\mathsf{C1}$};
   \node[main node][label={[xshift=0.65cm, yshift=-0.9cm]$\IN^*$}] (liable) [above  of=harm, fill = green!40] {$\mathsf{C2}$};
        
    \path[->,shorten >=1pt,auto, thick,every node/.style={font=\sffamily\small}]
    (guidelines) edge node {} (noGuidelines)
    (noGuidelines) edge node [right] {} (guidelines)

    (guidelines) edge node {} (noNegligent)
    (noGuidelines) edge node [right] {} (negligent)

    (negligent) edge node {} (noNegligent)
    (noNegligent) edge node [right] {} (negligent)
    
    (guidelines) edge node {} (noLiable)
    
    (negligent) edge node {} (noLiable)
    
    (noLiable) edge node [right] {} (liable);   
\end{tikzpicture}
\caption{Grounding of the BP-labelling with $BP(\{liable, \neg negligent\})$.
}
\labelfig{medical-example-no-burden}
\end{figure}

\end{example}

\begin{example}[Self-defeating argument]\label{example-self-defeating-def10}

Now let us consider a self-defeating argument. Let us assume the following premises with $\BP(\neg c)$:
\begin{center}
\begin{tabular}{@{}l l ll}
$\Rightarrow \neg c \quad$ &  $\neg c \Rightarrow p \quad$ & $ p \Rightarrow \ c$\\
\end{tabular}
\end{center}

\noindent We can build the following arguments:
\begin{center}
\begin{tabular}{ l l l l l}
$\mathsf{A1}:$ &$\Rightarrow \neg c$ \\ $\mathsf{A2}:$ &$\mathsf{A}1 \Rightarrow p$\\  $\mathsf{A3}:$ &$ \mathsf{A2} \Rightarrow c$\\
\end{tabular}
\end{center}

\noindent The  BP-labelling is depicted in \xf{example-self-defeat-fig}.
By Definition \ref{burdenLabelling} and item 3.(b)(i), argument $\mathsf{A1}$ is labelled $\OUT^*$, if its attacker $\mathsf{A3}$ is not $\OUT^*$.
By item 3.(b)(ii), argument $\mathsf{A2}$ is labelled $\OUT^*$.
By Definition \ref{burdenLabelling}, argument  $\mathsf{A3}$ cannot be labelled $\IN^*$, and thus it is labelled $\UND^*$. 
It turns out that in this case there is no complete BP-labelling, and thus no grounding.

\begin{figure}[!ht]
\centering
\begin{tikzpicture}[node distance=1.4cm,main node/.style={circle,fill=white!20,draw,font=\sffamily\scriptsize}]
    \node[main node][label={[xshift=-0.8cm, yshift=-0.9cm]$\OUT^*$}] (a1) [fill = red!40] {$\mathsf{A1}$};
    \node[main node][label={[xshift=-0.9cm, yshift=-0.9cm]$\OUT^*$}] (a2) [above of=a1, fill = red!40] {$\mathsf{A2}$};
    \node[main node][label={[xshift=-0.8cm, yshift=-0.9cm]$\UND^*$}] (a3) [above of=a2, fill = blue!40] {$\mathsf{A3}$};
    
    \path[->,shorten >=1pt,auto, thick,every node/.style={font=\sffamily\small}]
    (a1) edge[bend right=-30] node {} (a3)
    (a3) edge[bend right=-30] node {} (a1)    
    (a3) edge[bend right=-30] node {} (a2)
    (a3) edge[loop above] node {} (a3) ;   
    
\end{tikzpicture}
\caption{BP-labelling with $\BP(\neg c)$ where $\neg c = \Conc(\mathsf{A}1)$.}
\labelfig{example-self-defeat-fig}
\end{figure}
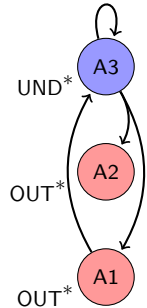

\end{example}



\section{Conclusion}\labelsec{section:conclusion}
A formal model of the burden of persuasion has been provided. The model shows how an allocation of the burden of persuasion may induce single outcomes ($\IN^*$ arguments), in contexts in which the assessment of conflicting arguments would otherwise remain undecided. 

Our model is a preliminary exploration of the intersection between the burden of persuasion and argumentation labelling frameworks, and it can provide a starting point for further research. In particular, it combines the insight of \cite{PrakkenSartor2008MP,PrakkenSartor2011MB}, where the burden of persuasion provides a criterion for adjudicating conflicts of arguments, and the insight of \cite{Gordon2007,GordonWalton2009PB}, where the satisfaction of burdens of argumentation depends on the dialectical status of the arguments concerned. The proposed model also deals with situations in which we have to combine a general burden of persuasion for one party, with exceptional propositions for which the burden is shifted onto the other party (so-called inversion of the burden of persuasion, see \cite{PrakkenSartor2011MB}.

The model can be extended to take into account different standards of proof that are required for meeting the burden of persuasion, such as the standard ``preponderance of the evidence'', which applies in private law, and the more rigorous ``beyond reasonable doubt'' standard of criminal law (see \cite{PrakkenSartor2011MB}).

\bibliographystyle{eptcs}
\bibliography{BP-ICLP2020}

\end{document}